\documentclass[sigconf, nonacm]{acmart}
\usepackage{hyperref}
\usepackage{url}
\usepackage{footnote}
\usepackage{booktabs}
\usepackage{threeparttable}

\newcommand{\tabincell}[2]{\begin{tabular}{@{}#1@{}}#2\end{tabular}}

\begin{document}
\title{A Note on General Statistics of Publicly Accessible\\ Knowledge Bases}

\author{Feixiang Wang, Yixiang Fang, Yan Song, Shuang Li, Xinyun Chen}
\affiliation{%
  \institution{The Chinese University of Hong Kong, Shenzhen}
}
\email{feixiangwang@link.cuhk.edu.cn, {fangyixiang, songyan, lishuang, chenxinyun}@cuhk.edu.cn}

\begin{abstract}
Knowledge bases are prevalent in various domains
and
have been widely used in a large number of real applications such as applications in online encyclopedia, social media, biomedical fields, bibliographical networks.
Due to their great importance, knowledge bases have received much attention from both the academia and industry community in recent years.
In this paper, we provide a summary of the general statistics of several open-source and publicly accessible knowledge bases, ranging from the number of objects, relations to the object types and the relation types.
With such statistics, this concise note can not only help researchers form a better and quick understanding of existing publicly accessible knowledge bases, but can also guide the general audience to use the resource effectively when they conduct research with knowledge bases.
\end{abstract}

\maketitle


\section{Introduction}

Knowledge bases are prevalent in various domains such as online encyclopedias (e.g., DBpedia), social media (e.g., Twitter), biomedical fields (e.g., Geonames), and bibliographical networks (e.g., DBLP).
Essentially, a knowledge base is defined as a collection of interlinked descriptions of entities (e.g., real-world objects and persons) that enables storage, analysis, and reuse of this knowledge in a machine-interpretable manner \cite{Knowledge_graph_definition}.
With the fact that the knowledge bases describe various properties and relationships between different facts, the complex relationships among different entities are thus well preserved and represented.

Nowadays, knowledge bases have been widely used in a large number of real applications.
For example, Google uses the knowledge graph to enhance its search engine.
Specifically, by using the knowledge graph, Google search engine can identify different meanings for the same word and summarize relevant content of the word to provide better search results (e.g., the word ``Taj Mahal'' may denote the monument or a musician, but Google search engine can show their difference when searching the relevant information like ``Taj Mahal's history''\footnote{\url{https://blog.google/products/search/introducing-knowledge-graph-things-not/}}).
For another example, the knowledge graph in Siemens has supported the knowledge-driven application across companies and generates new knowledge \cite{Siemens}.


Owing to their great importance, knowledge bases have received much attention from both the academia and industry community.
Very often, researchers need to resort to publicly accessible knowledge bases to evaluate their methods. Ideally, the knowledge bases need to offer insights to the devised models and solutions, and meanwhile should be user friendly and enable researchers to easily evaluate their methods.
%
In light of this, in this paper, we provide a note on general statistics of several publicly accessible knowledge bases, which include their numbers of objects, relations, object types, relation types, etc.
Table \ref{tab:general statistics} lists the publicly accessible knowledge bases that are reviewed in this paper\footnote{The statistics data were extracted from the official websites of the knowledge bases on June 28, 2021.}.
As shown in Table \ref{tab:general statistics}, we group these knowledge bases into two categories, i.e., schema-rich and simple-schema knowledge bases.
Herein, the schema of a knowledge base refers to a network describing all allowable relation types between entities types and also the properties of entities \cite{XML_schema}.
For example, if an entity is ``Taylor Swif'', then the schema defines the type of this entity (i.e., Person) and the properties of this entity in this type (i.e., her age, job, country, etc.).
Therefore, the schema-rich knowledge bases often refer to the knowledge bases with a large number of object/relation types, while the simple-schema knowledge bases only have a few object/relation types.
We would like to remark that apart from these, there are many other knowledge graphs whose entire data are not publicly available.
For example, for the enterprise knowledge bases (e.g., Twitter, Flickr, and Google), only a small proportion of their data are accessible.
For these knowledge bases, we will not cover in this paper.

Note that, Table \ref{tab:general statistics} only includes several publicly accessible knowledge bases according to the popularity in academic research.
We will continue to update this table by including more open-source knowledge bases in the future. 


\footnotetext[1]{\url{https://blog.google/products/search/introducing-knowledge-graph-things-not/}}

\begin{table*}[h]
\centering
\caption{Publicly accessible knowledge bases reviewed in this paper.}
\label{tab:general statistics}
\begin{tabular}{l c c c c c c}
\hline
Name&Schema type&URL&Major object types& Objects/entities&Initially released\\
\hline\hline
Yago&Schema-rich&https://yago-knowledge.org/&10K&67M&2008\\
\hline
DBpedia&Schema-rich&https://www.dbpedia.org/&484k&6 M&2007\\
\hline
Wikidata&Schema-rich&https://www.wikidata.org/&24M&93 M&2012\\
\hline
KBpedia&Schema-rich&https://kbpedia.org/&173 (ontology)&40M&2016\\
\hline
DBLP&Simple Schema&https://dblp.org/&4&8M&1993\\
\hline
IMDb&Simple Schema&https://www.imdb.com/&7&186M&1990\\
\hline
Geonames&Simple Schema&https://www.geonames.org/&9&12M&2005\\
\hline
\end{tabular}
\end{table*}


\section{General Statistics of Publicly Accessible Knowledge Bases}

We divide the general statistics of the publicly accessible knowledge bases into two parts.
The first part contains the common statistics of all knowledge bases, while the second part includes specific statistics.
For the first part, we mainly show the schema type, URL, initially released time, and the numbers of objects/entities, relations, and object types of each knowledge base. \autoref{tab:general statistics} summarizes the common statistics of these knowledge bases, where all the statistic numbers were collected from the official websites in June 2021 and they may change as time goes on.
Note that for some knowledge bases, the entity types are hierarchically organized, and here we mainly count the number of object types in high levels which is denoted by ``Major object types'' in \autoref{tab:general statistics}.
The second part covers the specific statistics for each knowledge base.
Since various knowledge bases have different structures, this leads to specified statistics.
Besides, the same concepts may have different meanings in each knowledge base.
For example, in \autoref{tab:general statistics}, the concept of entity types in Yago means the types for all the entities and ontology classes, while it refers to the type of entities for DBLP.
Therefore, apart from the common information, one needs to further elaborate on their specific statistics.

In the following, we mainly focus on introducing the specific statistics of each publicly accessible knowledge base.
Note that this paper will not extensively discuss the specific statistics of Wikidata, since many other knowledge bases (e.g., DBpedia, Yago, and KBpedia) are built based on this data.  

\subsection{Schema-rich Knowledge Bases}

In general, schema-rich knowledge bases refer to the knowledge bases consisting of entities with a large number of types. For example, Yago covers the information of various areas such as people, cities, countries, movies, and organizations \cite{Yago}, making it a knowledge base with various types of objects.

\subsubsection{DBpedia}
\indent\setlength{\parindent}{1em}DBpedia was created by scientists at the Free University of Berlin and Leipzig University in collaboration with OpenLink Software\cite{DBpedia_version_2007}. Now DBpedia is maintained by scientists at the University of Mannheim and Leipzig University.

\begin{figure}
  \centering
  \includegraphics[width=\linewidth]{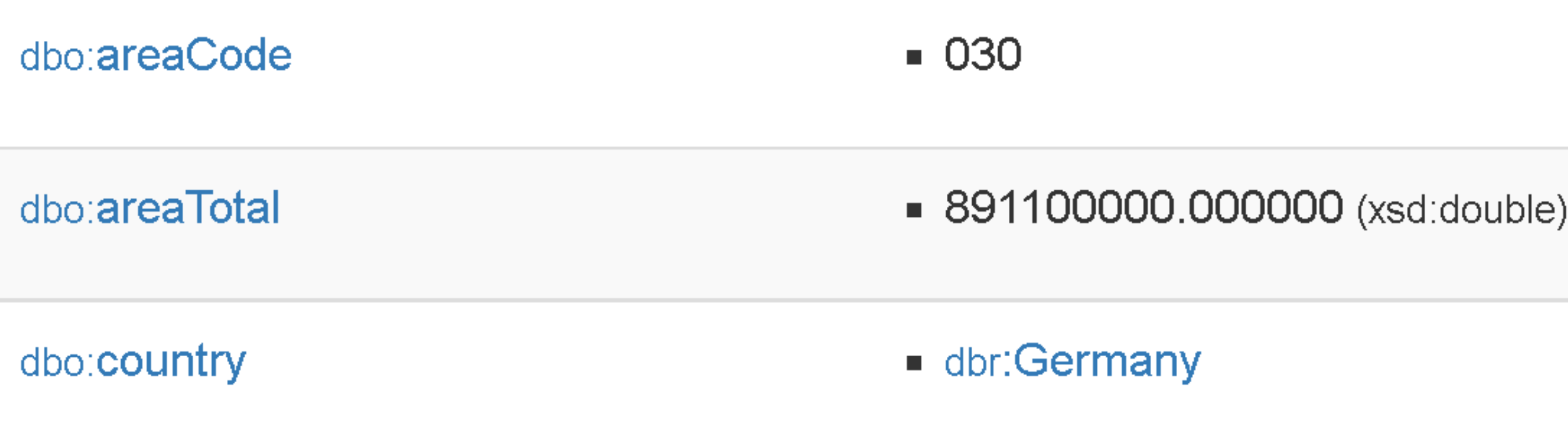}
  \caption{``Berlin'' recorded in DBpedia.}
  \label{fig:DBpedia dbo}
\end{figure}

DBpedia mines data from ``Wikipedia'', ``DBpedia common'',  and ``Wikidata'', and updates the data around the 15th of every month, allowing users to find answers to questions where the information is spread across multiple Wikipedia articles. For example, the word ``Berlin'' has many different meanings like Berlin, West-Berlin, and Ost-Berlin. By querying DBpedia with the keyword ``Berlin'', it will return descriptions about Berlin, West-Berlin, and Ost-Berlin, even if you only search Berlin. It will also return many properties and relationships to other entities, e.g., the information of universities and famous people in Berlin in history. In \autoref{fig:DBpedia dbo}\footnote{\url{https://dbpedia.org/page/Berlin}}, we choose some properties of Berlin and relationships to other entities. For example, ``Berlin'' has property ``areaCode'' and its value is ``030''. ``Berlin'' has a relationship to ``German'' and the relationship is ``country''.

The specific statistics of DBpedia are presented in \autoref{tab:DB specified statistics}\footnote{\url{https://www.dbpedia.org/blog/yeah-we-did-it-again-new-2016-04-dbpedia-release/}} and \autoref{tab:DB ontology}\footnote{\url{https://www.dbpedia.org/blog/yeah-we-did-it-again-new-2016-04-dbpedia-release/}}. Object property defines the relationship between one entity and another entity. Datatype property defines the relationship between one object and one of its values. For example, ``mother'' is an object property because mother is an individual as an entity, while ``age'' is a datatype property because a human's age is a value belonging to this human, not an entity. Languages symbolize the languages number knowledge base supports. Mappings define the mappings from Wikidata templates to the DBpedia classes. The mappings are classified by language. For example, in these 5800 mappings, there are 646 mappings are used in Dutch.
\autoref{tab:DB ontology} shows the distribution of ontology entities of DBpedia. Ontology classes mean the number of classes in ontology. Some classes are not in ontology but in knowledge bases. There are 5.2M entities in ontology classes. The classes having the largest size are Persons, Places, Works, Organizations, Species, and Diseases. In \autoref{tab:DB ontology}, we show the numbers of entities of each class having a large size.

\begin{table}[h]
    \centering
    \caption{Specific statistics of DBpedia.}
    \label{tab:DB specified statistics}
    \begin{tabular}{c c c c c }
        \hline
        \tabincell{c}{Object\\properties}
        &\tabincell{c}{Datatype\\properties}
        &Languages
        &Mappings\\
        \hline\hline
        1105&1754&136&5800\\
        \hline
    \end{tabular}
\end{table}

\begin{table}[ht]
    \caption{The distribution of ontology entity of DBpedia.}
    \label{tab:DB ontology}
    \begin{tabular}{l c}
        \hline
        Item&Number\\
        \hline\hline
        Ontology classes&754\\
        \hline
        Entities in ontology classes&5.2M\\
        \hline
        Persons&1.5M\\
        \hline
        Places&810k\\
        \hline
        Works&490k\\
        \hline
        Organizations&275k\\
        \hline
        Species&301k\\
        \hline
        Diseases&5k\\
        \hline
        Relationships/facts&9.5B\\
        \hline
    \end{tabular}
\end{table}

\subsubsection{Yago}
Yago is an open-source knowledge base developed at the Max Planck Institute for Computer Science in Saarbrücken. It is a knowledge graph containing knowledge about people, cities, countries, movies, and organizations. Each entity in Yago belongs to at least one class. The high classes in Yago are taken from schema.org and bioschemas.org, and the lower classes are a selection of classes from Wikidata. Yago has been used in the Watson artificial intelligence system \cite{Watson_artificial_intelligence}.

Yago extracted facts/relationships from Wikidata and Wikipedia, cleaned these data, and finally organized the cleaned data in the structures defined by the anthologies of ``Schema.org'' and ``Biosche-ma.org'', which define the number and the kind of classes, properties, and relationships.
The facts/relationships that are not fitted in these classes were ignored.
Yago used these classes as high-level classes and further built the low-level classes by extracting classes from Wikidata to make the whole ontology more clear.
If Wikipedia articles lack some entities' classes and properties, Yago will check the labels in Wikipedia. If Wikipedia cannot satisfy the need, Yago will further extract data from Wikidata \cite{Yago}.

\begin{figure}
  \centering
  \includegraphics[width=\linewidth]{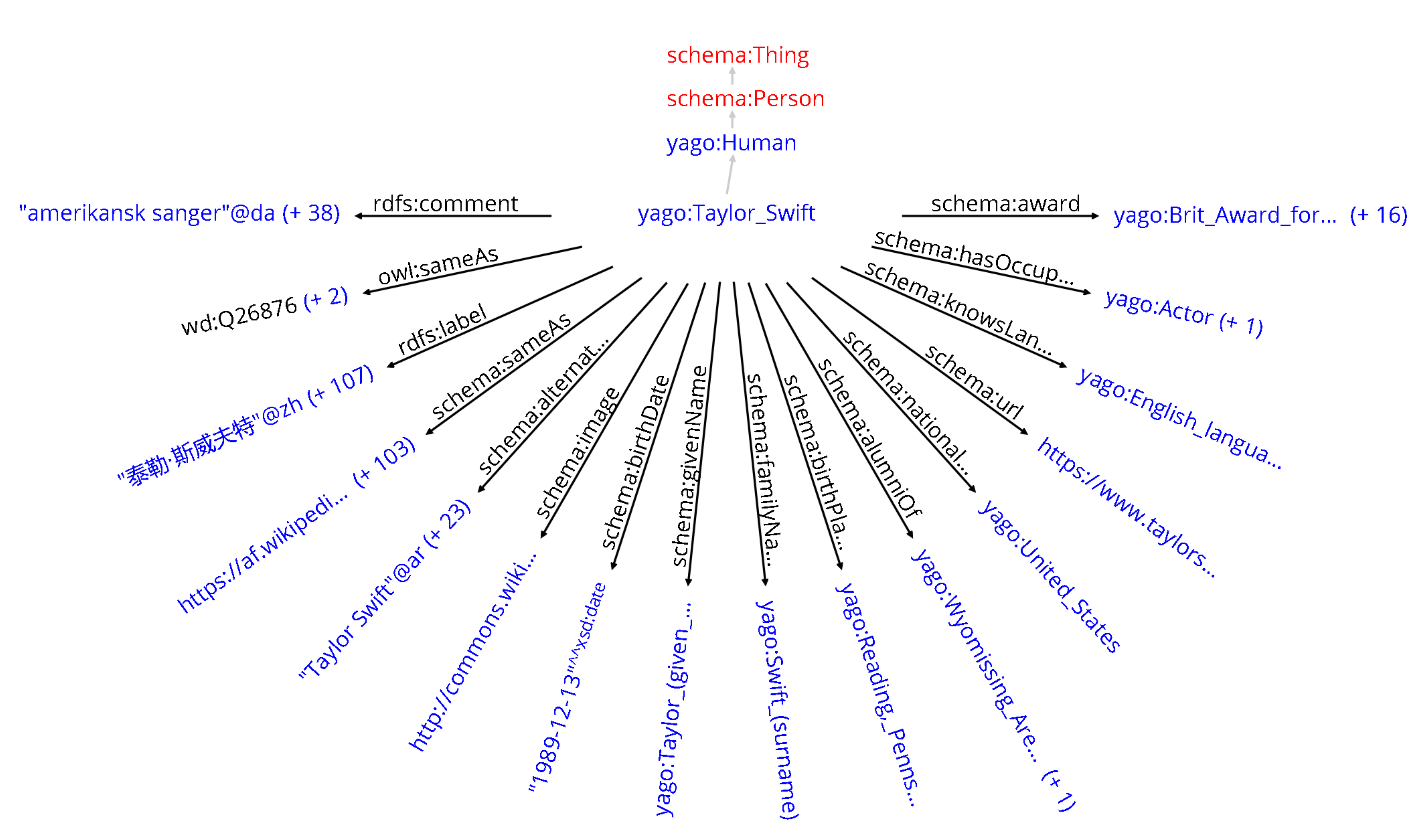}
  \caption{Using Yago to search ``Taylor Swift''.}
  \label{fig:Yago example}
 
\end{figure}

\begin{figure}
  \centering
  \includegraphics[width=\linewidth]{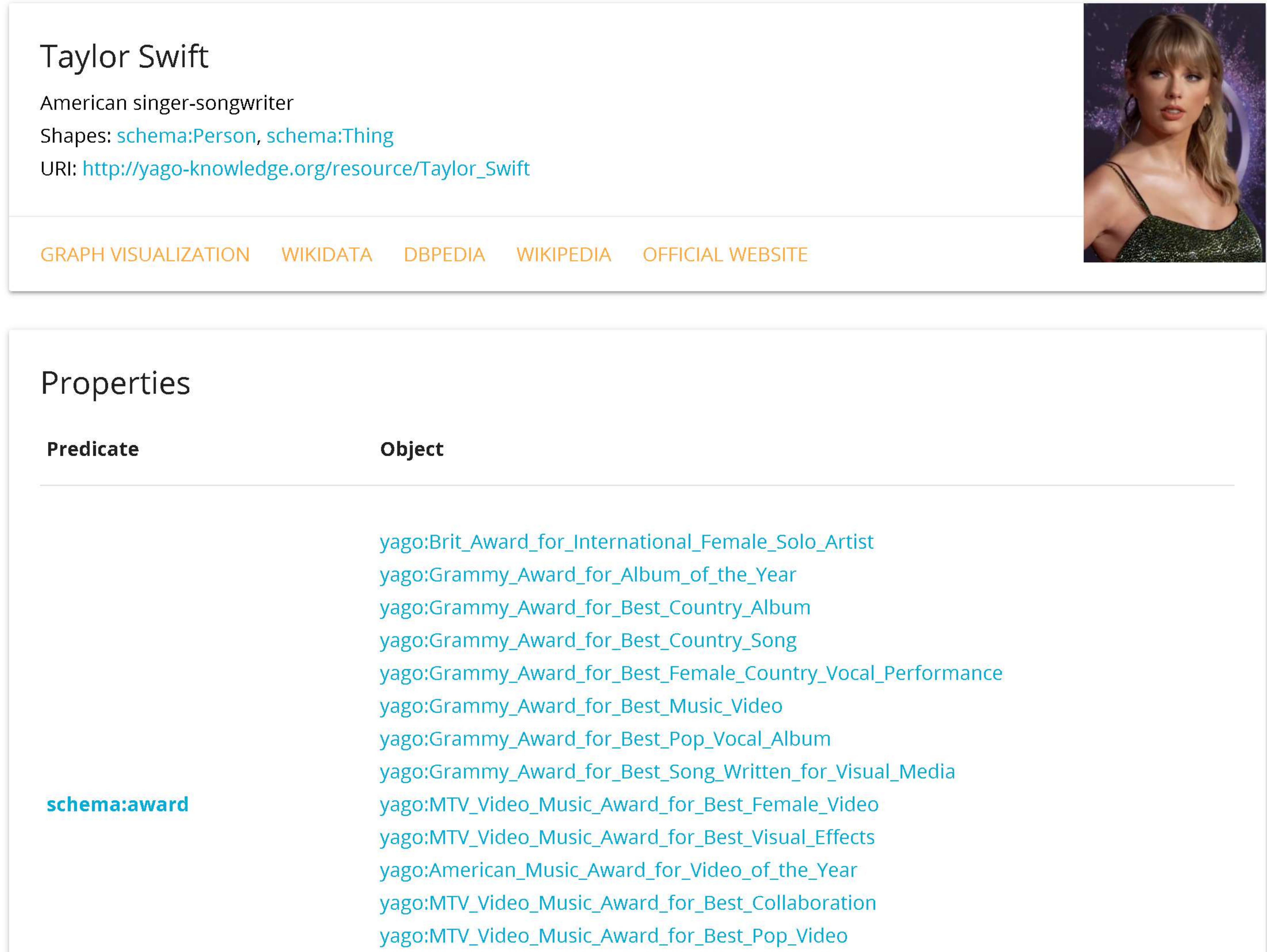}
  \caption{Using Yago to search ``Taylor Swift''.}
  \label{fig:Yago sepcified example}

\end{figure}

\autoref{fig:Yago example}\footnote{\url{https://yago-knowledge.org/graph/yago:Taylor_Swift}} shows an example of the information of the famous singer ``Taylor$\_$Swift'' in Yago. This graph only shows a few of relationships and properties belonging to ``Taylor$\_$Swift''. In terms of top classes, Taylor Swift is a singer, so it belongs to the class ``Human'', which is a subclass of class ``Person'' . Note that ``Person'' is a subclass of the highest classes ``Thing'' from Schema.org.
For properties of ``Taylor$\_$Swift'', the symbol ``(+x)'' after every properties or entities is the number of relationships in this relationship or property. In \autoref{fig:Yago example}, we see ``rdfs:label'' and ``rdfs:comment'', which are datatype properties of ``Taylor$\_$Swift''.
``Rdfs:label" means her different names in different languages, e.g., Chinese name.
In \autoref{fig:Yago example}, ``schema:X'' symbolizes a relationship between entity``Taylor$\_$Swift'' and other entities where this X refers to different contents, e.g.,  ``schema:nationality(schema:national...)'' shows that the country of ``Taylor$\_$Swift'' is another entity ``United States''.
``Owl:sameAs'' is a link to other knowledge bases. In this example, ``wd:Q26875'' symbolizes ``Taylor$\_$Swift'' in Wikidata, and ``(+2)'' means that there are 1+2=3 connections to other knowledge bases.
\autoref{fig:Yago example} only shows a part of relationships and properties about entity ``Taylor Swif''. \autoref{fig:Yago sepcified example} is the page showing all related information about entity ``Taylor Swift''.

\autoref{tab:Yago ontology}\footnote{\url{https://yago-knowledge.org/resource/yago:Taylor_Swift}} presents specific statistics of Yago. Here, the domain constraints mean some relationships only apply to some classes. For instance, ``birthPlace'' can only apply to a person and a place. Object range constraints and Datatype range constraints are similar to Domain constraints in function. Four of six major top-level classes (i.e., ``schema:BioChemicalEntity'', ``schema:Event'', ``schema:Person'', and ``schema:CreativeWork'') are pairwise disjoint, which implies that these classes cannot have any instances in common. As a result, the disjoint constraints can make these instances different\cite{Yago}. Note that these statistics are collected from the latest version of the Yago, called Yago 4, which is an RDFS knowledge base released in 2020.

\begin{table}[h]
\caption{Yago ontology statistics \cite{Yago}.}
\centering
\label{tab:Yago ontology}
\begin{tabular}{l c}
\hline
Item&Number\\
\hline\hline
Schema.org classes&235\\
\hline
Bioschema.org classes&6\\
\hline
Object Properties&100\\
\hline
Datatype properties&41\\
\hline
Node shapes&49\\
\hline
Property shapes&247\\
\hline
Domain constraints&247\\
\hline
Object range constraints&132\\
\hline
Datatype range constraints&57\\
\hline
Regex constraints&21\\
\hline
Disjoint constraints&18\\
\hline
Relationships/Facts&343M\\
\hline
\end{tabular}
\end{table}

\subsubsection{KBpedia}
The co-editors of KBpedia are Mike Bergman and Frédérick Giasson. Partial sponsorship of KBpedia has been provided by Cognonto Corporation. KBpedia was constructed by combining seven core knowledge bases, i.e., Wikipedia, Wikidata, schema.org, DBpedia, GeoNames, OpenCyc, and standard UNSPSC products and services. Until now, KBpedia has covered 98$\%$ Wikidata and nearly completely covers Wikipedia.

The upper structure of the KBpedia Knowledge Ontology (KKO) is informed by the triadic logic and universal categories of Charles Sanders Peirce. There are three categories in Peirce's trichotomy. The first category is about possibilities or potentials, the basic forces or qualities that combine or interact in various ways to enable the real things we perceive in the world, such as matter, life and ideas. The second category is about particular realized things or concepts we can describe such as an event, individual, entity. The third category is about laws, habits, regularities and continuities that may be generalized from particulars or possibilities. All generals — what are also known as classes, kinds or types — belong to this category\footnote{\url{https://kbpedia.org/docs/kko-upper-structure/}}.

Now, we use an example to introduce KBpedia structure. We use In KBpedia, each record has eight sections. The eight sections are ``Header'', ``Core Structure'', ``Extended Linkages'', ``Typologies'', ``Entities'', ``Aspect-related Entities'', ``Broader Concepts'', and``Narrower Concepts''. ``Header'' contains the alternative names, definition of ``Currency'' and image about this class (concept) if it has. The core structure contains sub-classes, super-classes and equivalent classes. Each item in this class is a mapping to the class in core knowledge bases. ``Extended Linkages'' is similar to ``Core structure'' but is connecting to more knowledge bases. Core knowledge bases are OpenCyc, UMBEL, GeoNames, DBpedia, Wikipedia and Wikidata. ``Entities'' is the same to the concept ``Entities'' in other knowledge bases. This item contains many entities and each entity has its properties and relationships with other entities. ``Aspect-related Entities'' are the entities related to class ``Currency'' but it provides a classifications for entities by situation for example related entities in financial. ``Typologies'' contain the core and extended typologies (classes) that the class ``Currency" belongs to. ``Typologies'' concept is similar to ``classes". KBpedia has 30 core typologies in total such as ``visual information''. Extended typologies indicate typologies that are not included in core typologies. ``Broader Concepts'' and ``Narrow Concepts'' correspond to super-classes and sub-classes.
\footnote{\url{https://kbpedia.org/use-cases/browse-the-knowledge-graph/}}

\autoref{tab:KBpedia statistics}\footnote{\url{https://kbpedia.org/resources/statistics/}} shows the ontology statistics of KBpedia. Ontology classes indicate the number of classes in ontology. Mapped vocabularies mean the number of mappings to those 7 knowledge bases and  16 extended vocabularies. Properties mean the number of relationships between different entities or entities and its properties.

\begin{table}[ht]
\caption{Some statistics of KBpedia ontology.}
\centering
\label{tab:KBpedia statistics}
\begin{tabular}{ll}
\hline
&Number\\
\hline
ontology classes&173\\
\hline
Number of mapped vocabularies&23\\
\hline
No. of properties(Sum)&5004\\
\hline
mappings to Wikidata&3970\\
\hline
mappings to schema.org&877\\
\hline
\end{tabular}
\end{table}

\subsection{Simple-schema Knowledge Bases}
Simple-schema knowledge bases are often designed for describing the information of objects in one particular area, so the numbers of object types and relation types are often very limited.

\subsubsection{DBLP}
As a computer science bibliography website, the DBLP was created in 1993 at Universität Trier in Germany. DBLP grew and became an organization hosting a programming bibliography database site from a collection of HTML files \cite{DBLP_lessons_learned}. DBLP became a branch of Schloss Dagstuhl – Leibniz-Zentrum für Informatik (LZI) in Novemberm 2018.

\begin{figure}[h]
  \centering
  \includegraphics[width=\linewidth]{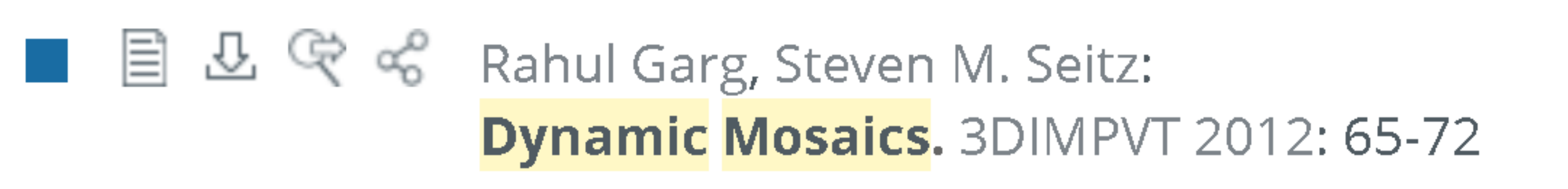}
  \caption{An example research paper in DBLP.}
  \label{fig:DBlP example}
\end{figure}

The typical types of objects in DBLP are publications, authors, conferences, journals, research topics, etc. \autoref{fig:DBlP example}\footnote{\url{https://dblp.org/search?q=hyperdoc2vec}} shows an example of a research paper record in DBLP, which includes five authors, paper title, conference name, published year, and page number.
\autoref{tab:DBLP general statistics}\footnote{\url{https://dblp.org/statistics/index.html}} and \autoref{tab:DBLP articles distribution}\footnote{\url{https://dblp.org/statistics/distributionofpublicationtype.html}} report some specific statistics of DBLP.\footnote{\url{https://dblp.org/statistics/distributionofpublicationtype.html}} \autoref{tab:DBLP general statistics} shows 4 top classes and the number of entities belong to each of them. For publications, DBLP has more detailed classification as shown in \autoref{tab:DBLP articles distribution}.

\begin{table}[ht]
\centering
\caption{The numbers of objects in each type in DBLP.}
\label{tab:DBLP general statistics}
\begin{tabular}{c c c c}
\hline       
Publications&Authors&Conferences&Journals\\
\hline\hline
5633863&2773696&5438&1764\\
\hline
\end{tabular}
\end{table}

\begin{table}[ht]
\centering
\caption{DBLP articles distribution.}
\label{tab:DBLP articles distribution}
\begin{tabular}{l r}
\hline
Type&Percentage\\
\hline\hline
References work&0.48$\%$\\
\hline
Books and thesis&1.74$\%$\\
\hline
Parts in Book or Collections&0.70$\%$\\
\hline
Journal articles&39.19$\%$\\
\hline
Informal publications&6.84$\%$\\
\hline
Editorship&0.88$\%$\\
\hline
Conference and workshop papers&50.17$\%$\\
\hline
\end{tabular}

\end{table}

In \autoref{fig:DBlP example}, we present an example of the article titled ``Dynamic Mosaics'', which was written by Rahul Garg and Steven M, Seitz. This article was published in conference 3DIMPVT 2012 and its page number is 65-72. Note that if this article was published in another venue like a journal, then the conference name would be replaced by the name of the venue.

\subsubsection{IMDb}
IMDb is an online database containing information related to films, television programs, home videos, video games, and streaming content online.
IMDb was established in 1990 as a movie database on the Usenet group ``rec.arts.movies''. It moved to the web in 1993. Now, IMDb is owned and operated by Amazon. 

\autoref{tab:IMDb statistics}\footnote{\url{https://www.imdb.com/pressroom/stats/}} shows the specific statistics for IMDb. IMDb mainly consist of seven types of entities, i.e., titles,
genres, filmography, names, images, and videos.
More precisely, the titles contain plot outline, literature, and keyword;
The genres contain drama, talk show, and romance;
The names contain article, birth name, etc.;
The filmography contains different people in different works such as actors, actresses, producers, writers, etc.; 
Images contain images, publicity images, posters, etc.;
Videos contain trailers, clips, feature films, etc.

\begin{figure}[h]
  \centering
  \includegraphics[width=\linewidth]{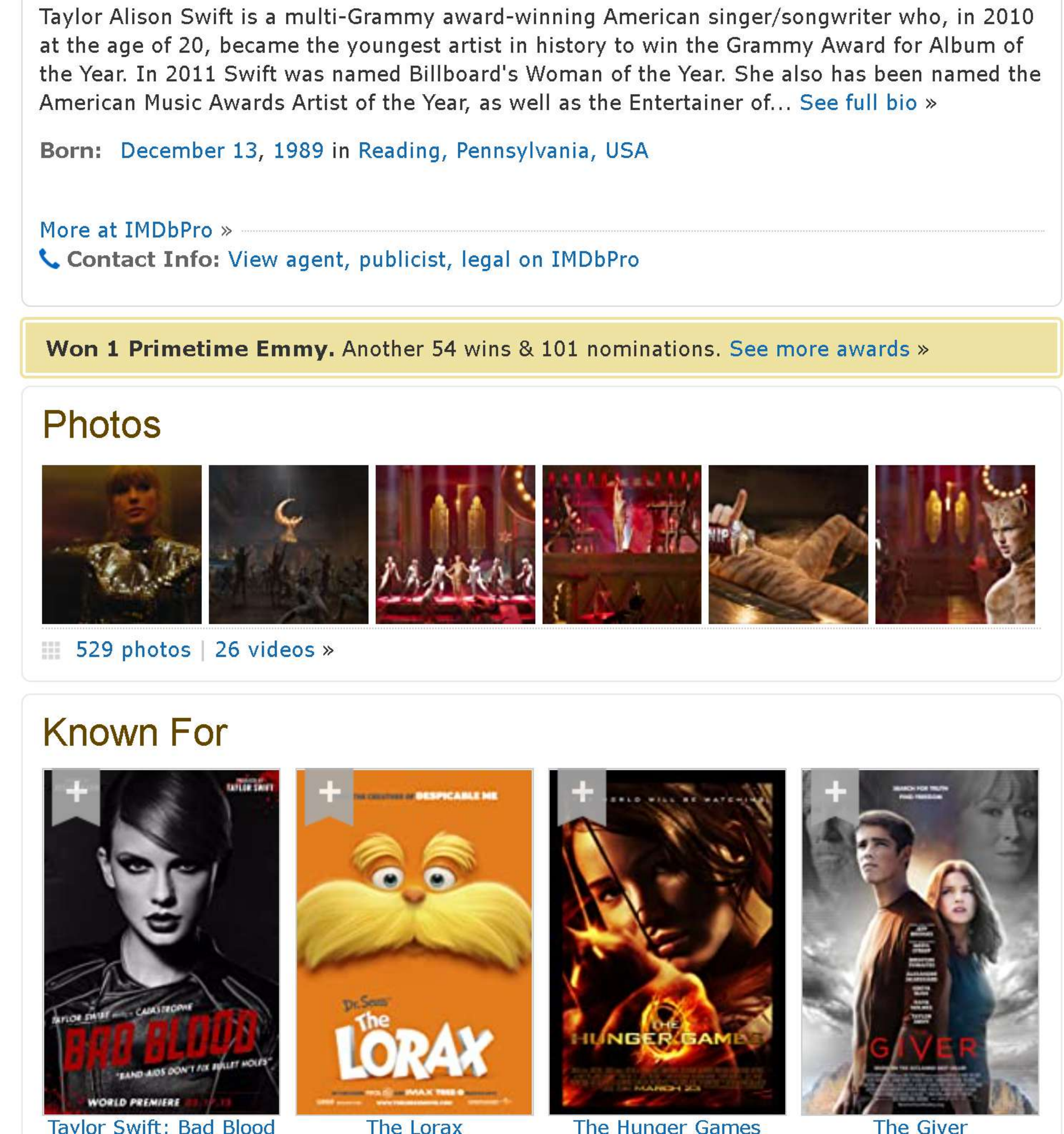}
  \caption{An example of search result in IMDb.}
  \label{fig:IM}
\end{figure}

To illustrate the data, we show a part of the information for the record of the singer ``Taylor Swift'' in \autoref{fig:IM}\footnote{\url{https://www.imdb.com/find?q=Taylor+Swift&ref_=nv_sr_sm}}. Here, ``Taylor Swift'' is an entity, and the introduction of ``Taylor Swift'' is a property of ``Taylor Swift''.
The relationship could be considered as ``rdfs:comments'' in Yago. Entity ``Taylor Swift'' has other properties and relationships.
In the complete record of ``Taylor Swift'', the entity has three relationships to other entities, i.e., ``Photos'', ``Known For'', ``Filmography''. For example, each photo is an entity and every single photo has properties like titles or some other properties.

\begin{table}[h]
\centering
\caption{Some specific statistics of IMDb.}
\label{tab:IMDb statistics}
\begin{tabular}{l c c}
\hline
Item&Number&subclass types\\
\hline
Titles&7766299&19\\
\hline
Genres&12951717&28\\
\hline
Filmography&142530261&32\\
\hline
Names&10827285&12\\
\hline
Images&11856447&8\\
\hline
Videos&471449&7\\
\hline
\end{tabular}

\end{table}

\subsubsection{GeoNames}
GeoNames is a user-editable geographical knowledge graph founded in 2005. GeoNames contains over 25,000,000 geographical names corresponding to over 11,800,000 unique features. Geonames is editable for everyone. In other words, if some false information entered the knowledge base and the information is not accessed frequently, it is difficult to find and fix it\cite{Geonames_accuracy}, which is one of the drawbacks of Geonames.
However, according to some studies, there is a way to adjust such information in the Geonames \cite{Geonames_improving}. 

Geonames has nine feature classes and 645 feature codes. All coordinates in Geonames use the World Geodetic System 1984 (WGS84). Each feature in Geonames has its corresponding URI, which provides access to HTML wiki page, or a RDF description of the feature through content negotiation, using elements of the GeoNames ontology. This ontology describes the GeoNames features properties using the Web Ontology Language (OWL), the feature classes and codes being described in the SKOS language. GeoNames data are linked to DBpedia data and other RDF Linked Data through Wikipedia url in RDF descriptions.

\begin{figure}
    \centering
    \includegraphics[width=\linewidth]{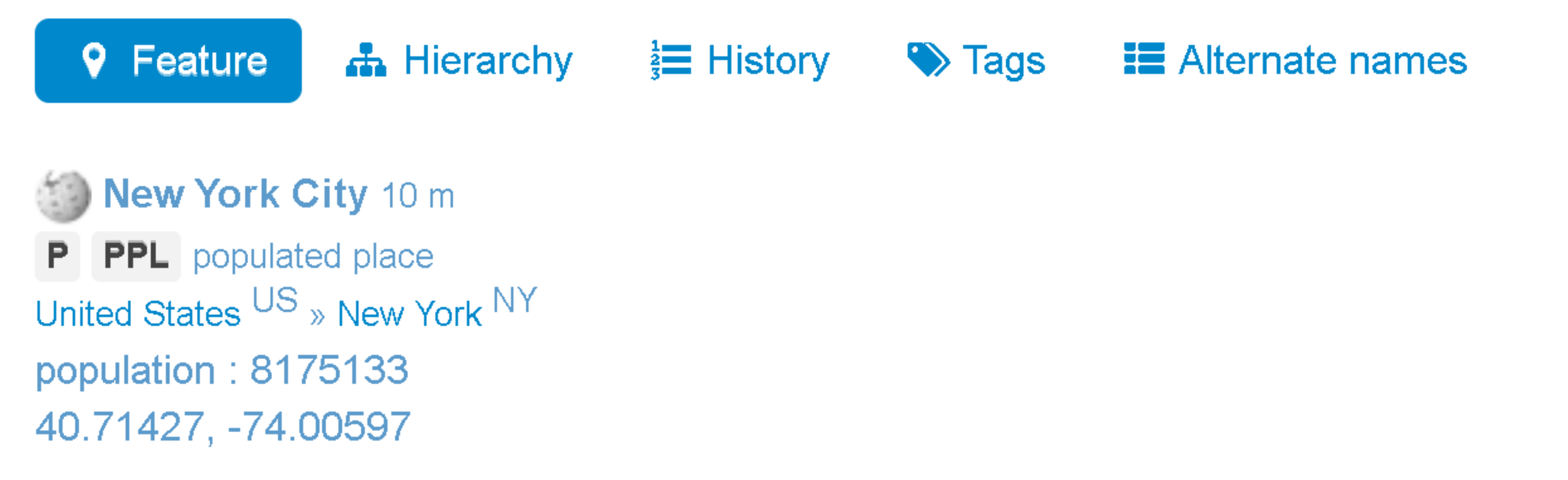}
    \caption{Example of ``New York City'' in GeoNames.}
    \label{fig:geo}
\end{figure}

We use the entity of ``New York City'' as an example to show the data of GeoNames. As shown in \autoref{fig:geo}\footnote{\url{https://www.geonames.org/5128581/new-york-city.html}}, the population is a property of ``New York City'' and its value is 8175133. Its first order administrative division is ``New York'' and its country is ``United States'', which are entities in GeoNames.

\autoref{tab:Geonames statistics}\footnote{\url{https://www.geonames.org/statistics/total.html}} presents some specific statistics data for GeoNames. Specifically, Geonames has 9 top classes \cite{schema_for_Geonames}:
Administrative Boundary Features describes the ways as how to define a complete Administrative boundary and its levels;
Hydrographic Features define the classes of hydrographic items like ``lake'', ``well'', and ``stream'';
Area Features contain ``park'', ``locality'', etc.;
Populated Place Features describe the features in area with many people such as populated locality and religious populated place;
Road/Railroad Features contain ``road junction'';
Spot features contain specified spot like ``farm'' and ``school'';
Hypsographic Features contain ``mountain'', ``hill'', ``island'', etc.;
Undersea Features contains ``shoals'', ``canyon'', etc.;
Vegetation Features contains ``forest'', ``grove'', etc.
A more detailed introduction of these statistics could be found from \cite{schema_for_Geonames}.

\begin{table}[h]
\centering
\caption{Some specific statistics of GeoNames.}
\label{tab:Geonames statistics}
\begin{tabular}{l c c}
\hline
Item&Number&Subfeatures Types\\
\hline\hline
Administrative Boundary Features&452363&25\\
\hline
Hydrographic Features&2236741&135\\
\hline
Area Features&417387&49\\
\hline
Populated Place Features&4793777&19\\
\hline
Road/Railroad Features&52112&21\\
\hline
Spot Features&2440626&254\\
\hline
Hypsographic Features&1598944&99\\
\hline
Undersea Features&14899&63\\
\hline
Vegetation Features&48709&18\\
\hline
Undefined&5058&\\
\hline
\end{tabular}
\end{table}

\section{Conclusions and Future Work}
In this paper, we present the general statistics of several knowledge bases, which are publicly accessible and receive much attention from both academic area and industry area. In particular, we divide the knowledge graphs into schema-rich knowledge bases and simple-schema knowledge bases. We introduce their common statistics and also the specific statistics of each knowledge base. We hope that by providing such statistics, the paper can make researchers have a better understanding of existing open-source knowledge bases.

In the future, we will continue to update the statistics of popular publicly  accessible knowledge bases. On the one hand, the existing knowledge bases are growing as new information is included. On the other hand, more knowledge bases will be released and made publicly accessible. Thus, we will keep updating this paper continuously to ensure the statistics data in this paper could provide up-to-date information.

\bibliographystyle{ACM-Reference-Format}
\bibliography{1,2,3,4,5,6,7,8,9,10,11}


\begin{thebibliography}{10}


\ifx \showCODEN    \undefined \def \showCODEN     #1{\unskip}     \fi
\ifx \showDOI      \undefined \def \showDOI       #1{#1}\fi
\ifx \showISBNx    \undefined \def \showISBNx     #1{\unskip}     \fi
\ifx \showISBNxiii \undefined \def \showISBNxiii  #1{\unskip}     \fi
\ifx \showISSN     \undefined \def \showISSN      #1{\unskip}     \fi
\ifx \showLCCN     \undefined \def \showLCCN      #1{\unskip}     \fi
\ifx \shownote     \undefined \def \shownote      #1{#1}          \fi
\ifx \showarticletitle \undefined \def \showarticletitle #1{#1}   \fi
\ifx \showURL      \undefined \def \showURL       {\relax}        \fi
\providecommand\bibfield[2]{#2}
\providecommand\bibinfo[2]{#2}
\providecommand\natexlab[1]{#1}
\providecommand\showeprint[2][]{arXiv:#2}

\bibitem[\protect\citeauthoryear{Ahlers}{Ahlers}{2013}]%
        {Geonames_accuracy}
\bibfield{author}{\bibinfo{person}{Dirk Ahlers}.}
  \bibinfo{year}{2013}\natexlab{}.
\newblock \showarticletitle{Assessment of the Accuracy of GeoNames Gazetteer
  Data}. In \bibinfo{booktitle}{\emph{Proceedings of the 7th Workshop on
  Geographic Information Retrieval}} (Orlando, Florida)
  \emph{(\bibinfo{series}{GIR '13})}. \bibinfo{publisher}{Association for
  Computing Machinery}, \bibinfo{address}{New York, NY, USA},
  \bibinfo{pages}{74–81}.
\newblock
\showISBNx{9781450322416}
\urldef\tempurl%
\url{https://doi.org/10.1145/2533888.2533938}
\showDOI{\tempurl}


\bibitem[\protect\citeauthoryear{Auer, Bizer, Kobilarov, Lehmann, Cyganiak, and
  Ives}{Auer et~al\mbox{.}}{2007}]%
        {DBpedia_version_2007}
\bibfield{author}{\bibinfo{person}{S{\"o}ren Auer}, \bibinfo{person}{Christian
  Bizer}, \bibinfo{person}{Georgi Kobilarov}, \bibinfo{person}{Jens Lehmann},
  \bibinfo{person}{Richard Cyganiak}, {and} \bibinfo{person}{Zachary Ives}.}
  \bibinfo{year}{2007}\natexlab{}.
\newblock \showarticletitle{Dbpedia: A nucleus for a web of open data}.
\newblock In \bibinfo{booktitle}{\emph{The semantic web}}.
  \bibinfo{publisher}{Springer}, \bibinfo{pages}{722--735}.
\newblock


\bibitem[\protect\citeauthoryear{Ehrlinger and W{\"o}{\ss}}{Ehrlinger and
  W{\"o}{\ss}}{2016}]%
        {Knowledge_graph_definition}
\bibfield{author}{\bibinfo{person}{Lisa Ehrlinger} {and}
  \bibinfo{person}{Wolfram W{\"o}{\ss}}.} \bibinfo{year}{2016}\natexlab{}.
\newblock \showarticletitle{Towards a Definition of Knowledge Graphs.}
\newblock \bibinfo{journal}{\emph{SEMANTiCS (Posters, Demos, SuCCESS)}}
  \bibinfo{volume}{48} (\bibinfo{year}{2016}), \bibinfo{pages}{1--4}.
\newblock


\bibitem[\protect\citeauthoryear{Ferrucci, Brown, Chu-Carroll, Fan, Gondek,
  Kalyanpur, Lally, Murdock, Nyberg, Prager, et~al\mbox{.}}{Ferrucci
  et~al\mbox{.}}{2010}]%
        {Watson_artificial_intelligence}
\bibfield{author}{\bibinfo{person}{David Ferrucci}, \bibinfo{person}{Eric
  Brown}, \bibinfo{person}{Jennifer Chu-Carroll}, \bibinfo{person}{James Fan},
  \bibinfo{person}{David Gondek}, \bibinfo{person}{Aditya~A Kalyanpur},
  \bibinfo{person}{Adam Lally}, \bibinfo{person}{J~William Murdock},
  \bibinfo{person}{Eric Nyberg}, \bibinfo{person}{John Prager},
  {et~al\mbox{.}}} \bibinfo{year}{2010}\natexlab{}.
\newblock \showarticletitle{Building Watson: An overview of the DeepQA
  project}.
\newblock \bibinfo{journal}{\emph{AI magazine}} \bibinfo{volume}{31},
  \bibinfo{number}{3} (\bibinfo{year}{2010}), \bibinfo{pages}{59--79}.
\newblock


\bibitem[\protect\citeauthoryear{Hubauer, Lamparter, Haase, and Herzig}{Hubauer
  et~al\mbox{.}}{2018}]%
        {Siemens}
\bibfield{author}{\bibinfo{person}{Thomas Hubauer}, \bibinfo{person}{Steffen
  Lamparter}, \bibinfo{person}{Peter Haase}, {and}
  \bibinfo{person}{Daniel~Markus Herzig}.} \bibinfo{year}{2018}\natexlab{}.
\newblock \showarticletitle{Use Cases of the Industrial Knowledge Graph at
  Siemens.}. In \bibinfo{booktitle}{\emph{International Semantic Web Conference
  (P\&D/Industry/BlueSky)}}.
\newblock


\bibitem[\protect\citeauthoryear{Ley}{Ley}{2009}]%
        {DBLP_lessons_learned}
\bibfield{author}{\bibinfo{person}{Michael Ley}.}
  \bibinfo{year}{2009}\natexlab{}.
\newblock \showarticletitle{DBLP: Some Lessons Learned}.
\newblock \bibinfo{journal}{\emph{Proc. VLDB Endow.}} \bibinfo{volume}{2},
  \bibinfo{number}{2} (\bibinfo{date}{Aug.} \bibinfo{year}{2009}),
  \bibinfo{pages}{1493–1500}.
\newblock
\showISSN{2150-8097}
\urldef\tempurl%
\url{https://doi.org/10.14778/1687553.1687577}
\showDOI{\tempurl}


\bibitem[\protect\citeauthoryear{Maltese and Farazi}{Maltese and
  Farazi}{2013}]%
        {schema_for_Geonames}
\bibfield{author}{\bibinfo{person}{Vincenzo Maltese} {and}
  \bibinfo{person}{Feroz Farazi}.} \bibinfo{year}{2013}\natexlab{}.
\newblock \showarticletitle{A semantic schema for GeoNames}.
\newblock  (\bibinfo{year}{2013}).
\newblock


\bibitem[\protect\citeauthoryear{Singh and Rafiei}{Singh and Rafiei}{2018}]%
        {Geonames_improving}
\bibfield{author}{\bibinfo{person}{Sanket~Kumar Singh} {and}
  \bibinfo{person}{Davood Rafiei}.} \bibinfo{year}{2018}\natexlab{}.
\newblock \showarticletitle{Strategies for geographical scoping and improving a
  gazetteer}. In \bibinfo{booktitle}{\emph{Proceedings of the 2018 World Wide
  Web Conference}}. \bibinfo{pages}{1663--1672}.
\newblock


\bibitem[\protect\citeauthoryear{Tanon, Weikum, and Suchanek}{Tanon
  et~al\mbox{.}}{2020}]%
        {Yago}
\bibfield{author}{\bibinfo{person}{Thomas~Pellissier Tanon},
  \bibinfo{person}{Gerhard Weikum}, {and} \bibinfo{person}{Fabian Suchanek}.}
  \bibinfo{year}{2020}\natexlab{}.
\newblock \showarticletitle{Yago 4: A reason-able knowledge base}. In
  \bibinfo{booktitle}{\emph{European Semantic Web Conference}}. Springer,
  \bibinfo{pages}{583--596}.
\newblock


\bibitem[\protect\citeauthoryear{Thompson, Beech, Maloney, Mendelsohn,
  Consortium, et~al\mbox{.}}{Thompson et~al\mbox{.}}{2001}]%
        {XML_schema}
\bibfield{author}{\bibinfo{person}{Henry~S Thompson}, \bibinfo{person}{David
  Beech}, \bibinfo{person}{Murray Maloney}, \bibinfo{person}{Noah Mendelsohn},
  \bibinfo{person}{World Wide~Web Consortium}, {et~al\mbox{.}}}
  \bibinfo{year}{2001}\natexlab{}.
\newblock \bibinfo{title}{XML schema part 1: Structures}.
\newblock
\newblock


\end{thebibliography}

\end{document}